\documentclass[prl,aps,amsfonts,amsmath,superscriptaddress,%
twocolumn,showpacs,floatfix]{revtex4}

\usepackage{amssymb}
\usepackage[dvips]{graphicx}
\usepackage[english]{babel}
\usepackage{indentfirst}
\usepackage{amsxtra}
\usepackage{amsmath}
\usepackage{multirow}
\usepackage [mathcal]{eucal}

\newcommand{\br}{{\bf r}}
\newcommand{\bk}{{\bf k}}
\newcommand{\bq}{{\bf q}}

\begin{document}
\title {\bf Beyond-mean-field theories with zero-range effective 
interactions. A way to handle the ultraviolet divergence}

\author{K. Moghrabi}
\affiliation{Institut de Physique Nucl\'eaire, IN2P3-CNRS, Univ. Paris-Sud, 
F-91406 Orsay Cedex, France}
\affiliation{Universit\'e Claude Benard Lyon 1, \'Ecole Normale Sup\'erieure de Lyon, 69364 Lyon Cedex, France}

\author{M. Grasso}
\affiliation{Institut de Physique Nucl\'eaire, IN2P3-CNRS, Univ. Paris-Sud, 
F-91406 Orsay Cedex, France}

\author{G. Col\`o}
\affiliation{Dipartimento di Fisica, Universit\`a  
degli Studi di Milano and INFN, Sezione di Milano, 20133 Milano, Italy}

\author{N. Van Giai}
\affiliation{Institut de Physique Nucl\'eaire, IN2P3-CNRS, Univ. Paris-Sud, 
F-91406 Orsay Cedex, France}

\begin{abstract}

Zero-range effective interactions are commonly used 
in nuclear physics and in other domains  
to describe many-body systems  
within the mean-field model. 
If they are used  
within a
beyond-mean-field framework, 
contributions to the total energy   
that display an ultraviolet divergence are found. 
We propose a general strategy to regularize this divergence 
and we illustrate it in the case of the second-order corrections to
the equation of state (EOS) of uniform symmetric matter. 
 By setting a  
momentum cutoff $\Lambda$, we show that  
for every (physically meaningful) value of $\Lambda$ it is possible to 
determine a new 
interaction such that the EOS 
with the second-order
corrections reproduces the empirical EOS, with a fit of the same quality as that obtained at the mean-field level. 

\end{abstract} 


 \maketitle



In many-body systems 
contact interactions are reasonable approximations to the realistic 
finite-range forces that can be employed in cases where the interaction range 
is smaller than the typical length scale represented by the interparticle distance. 
The main advantage of using 
zero-range interactions 
is that the   
equations to handle are 
greatly simplified. 
Two examples of commonly 
used contact interactions are 
the Skyrme forces which are quite popular  
in nuclear physics 
\cite{skyrme}, and the contact interactions with coupling 
strengths depending on the $s$-wave scattering length which are 
employed for dilute atomic gases 
(see, e.g., Ref. \cite{bruun}). 
In general, effective interactions 
contain
parameters that must be  
adjusted 
to reproduce a given set of observables (for instance,  
in the nuclear case, binding energies and radii of a  
few selected nuclei, and bulk   
properties of nuclear matter).
Usually this is done at the self-consistent  
mean-field level. When going beyond 
this level, 
it is likely that the effective interaction has to be redetermined. 
However, trying to discuss this issue is impossible in the case 
of contact interactions, since going beyond mean field, e.g., 
by including second-order corrections, implies
dealing with contributions to the total energy which contain  
an ultraviolet divergence. 

Many authors 
have studied the perturbation series for
the energy of the electron gas \cite{gellmann,sawada,nozieres} 
and of nuclear systems \cite{euler,huby,thouless,levinger}. 
In all cases, the interaction is finite-range and no ultraviolet divergence 
appears. 
In this paper, we apply to a simple case a strategy to 
handle the contact interactions up to second order. 
We include a momentum cutoff $\Lambda$ among the parameters
of the interaction, and we show that for every value of $\Lambda$ the 
other parameters can be determined in such a way that the 
total energy of the system with second-order contributions remains the same. 
This strategy, and the formulas we will show, are quite general and
can be applied to different 
Fermi systems.
For the numerical application we restrict ourselves to
symmetric nuclear matter, treated with a simplified zero-range interaction.

\begin{figure}[htb]
\begin{center}
\includegraphics[width=5cm]{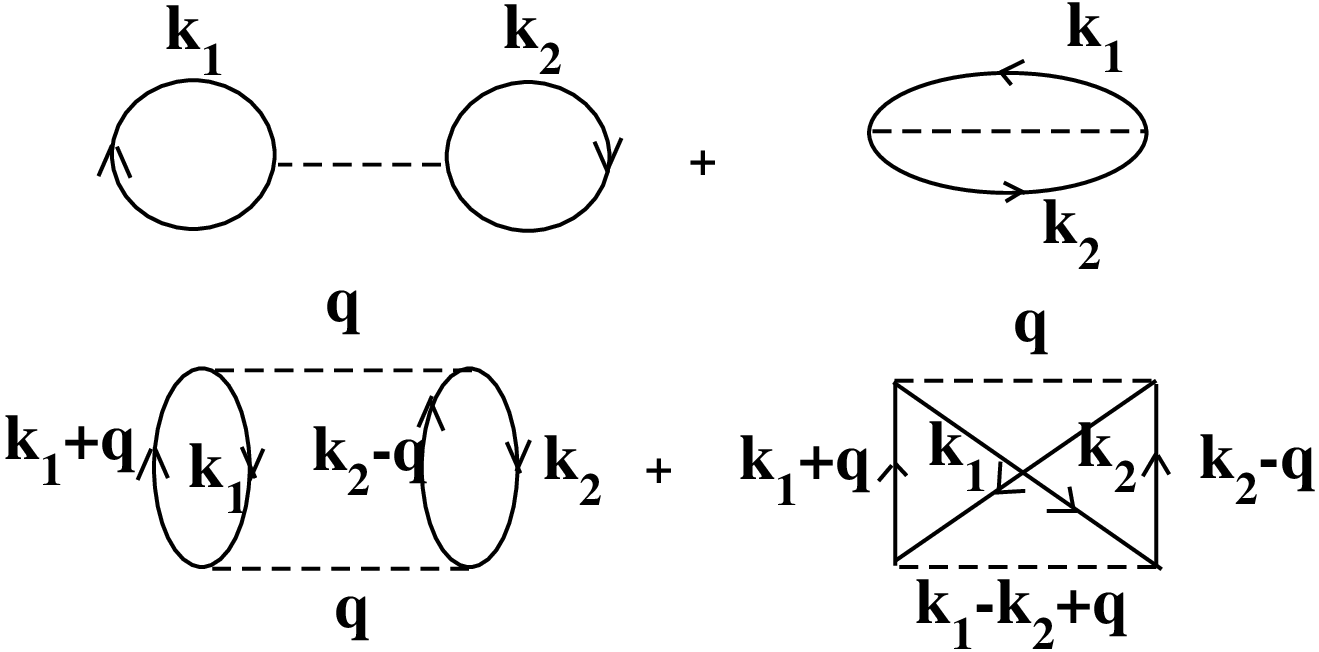}
\end{center}
\caption{\small First- and second-order diagrams for the total energy
in uniform 
matter. Labels refer to momentum states.}
\label{figdiag1}
\end{figure}

The  
second-order terms that 
contribute to
the total energy in uniform matter and diverge in the case of 
contact forces are
 shown in the lower part of Fig. \ref{figdiag1}. 
The divergence is caused
by the integration on $\bq$ and is somehow 
unphysical since the high-momentum states are certainly
outside the scale at which effective forces are to be used. 

In the case of effective interactions between point-nucleons, 
the cutoff $\Lambda$ must certainly be smaller than the momentum associated 
with the nucleon size, i.e., smaller than $\approx$ 2 fm$^{-1}$. 
In fact, these interactions are used to describe giant resonances or
rotational bands of 
nuclei and consequently the scale should
be even smaller, perhaps around 0.5 fm$^{-1}$. However, our
procedure is tailored on the basic idea of renormalizable
quantum-field theories: so, in principle, 
for any value of $\Lambda$
a new set of parameters 
is found which leads to the same   
equation of state (EOS) -including the second-order contributions-. 
We show below that this is mathematically possible 
owing to the fact that the 
second-order correction is 
well-behaved as a function of the 
density. 
Thus, we imitate the QED idea that the bare electron mass and charge
can be chosen for every different energy cutoff in such a way that
the physical values of mass and charge are always obtained.

An ultraviolet divergence appears already at the mean-field level in the  
Hartree-Fock-Bogoliubov (HFB) \cite{RS} or Bogoliubov-de Gennes 
(BdG) \cite{degennes}
models when zero-range forces are employed in the pairing channel. 
In this case, sophisticated regularization schemes exist 
in which the irregular term of the pairing field is suppressed 
and the dependence on the energy cutoff is eliminated. 
These techniques are commonly adopted 
for example in atomic physics in BdG models
\cite{bruun,grasso}, and have been also employed in 
nuclear physics \cite{bulgac},  
but they cannot be directly applied to the case of interest studied in 
the present work.



Let us  
write the zero-range force as
\begin{eqnarray}
V(\br_1,\br_2)= g \delta(\br_1-\br_2). 
\label{eq1}
\end{eqnarray}
To make contact with the Skyrme interactions \cite{skyrme} 
the strength $g$ is written as 
$t_0 + \frac{1}{6}t_3 \rho^{\alpha}$ and 
this 
corresponds to the so-called
$(t_0,t_3)$ model that is a simplification of the usual Skyrme model where  
the spin-dependent, velocity-dependent and 
spin-orbit terms    
are dropped. 
If the Skyrme force is viewed as a 
$G$ matrix (thus 
including ladder diagrams), the introduction of second-order contributions 
would in principle imply a double counting. However, our attitude is to consider as a matter of 
fact our interaction as phenomenological, and our framework as an effective theory where 
parameters are readjusted according to which diagrams are explicitly introduced. 

Normalizing the single-particle wave functions 
within a box of volume 
$\Omega$, 
the Hartree-Fock (HF) potential-energy contribution (upper part of 
Fig. \ref{figdiag1}) is equal to
\begin{eqnarray}
E=d \frac{\Omega^2}{(2 \pi)^6} \int_{k_1,k_2<k_F} d^3k_1 d^3k_2 
v(\bk_1,\bk_2,\bk_1,\bk_2),
\label{energy}
\end{eqnarray}
where $d=(n^2-n)/2$, 
$n$ being the level degeneracy (4 in the
case of symmetric nuclear matter), and
$v=\frac{g}{\Omega}$.
The energy per particle or EOS is obtained by
adding the kinetic contribution and reads in our case 
\begin{eqnarray}
\frac{E}{A} (\rho)= \frac{3\hbar^2}{10m}
\left(\frac{3\pi^2}{2}\rho\right)^{2/3} + \frac{3}{8}t_0\rho +
\frac{1}{16}t_3 \rho^{\alpha+1}.
\label{eq5}
\end{eqnarray}
Eq. (\ref{eq5}) 
coincides with the EOS obtained with the SkP 
parameter set \cite{skp}, in which no contribution coming from the 
velocity-dependent terms 
appears and the effective mass coincides with the bare mass. 
Since SkP, as all the Skyrme sets, has been fitted to reproduce
within HF 
the basic features of the 
nuclear 
EOS, one can consider its associated
energy per particle as a benchmark which must be reproduced with
reasonable accuracy for every value of $\Lambda$ when the
second-order correction is included.

This second-order correction 
is given by
\begin{eqnarray}
&& \Delta E = d \frac{\Omega^3}{(2 \pi)^9} \int_{k_1,k_2<k_F,|\bk_1+\bq|,
\bk_2-\bq|>k_F} d^3k_1 d^3k_2 d^3q \nonumber \\
&& \frac{v^2}{\epsilon_{\bk_1} + \epsilon_{\bk_2} -\epsilon_{\bk_1+\bq}-
\epsilon_{\bk_2-\bq}} 
\equiv C \int dq v^2 G(q), 
\label{eq6}
\end{eqnarray}
where $C=-6 m \Omega^3 (2 \pi)^{-9} \hbar^{-2}$ for symmetric nuclear matter, and 
$G(q)$ reads
\begin{eqnarray}
G(q)=\int_{k_1,k_2<k_F,|\bk_1+\bq|,\bk_2-\bq|>k_F}  
\frac{d^3k_1 d^3k_2}{q^2+\bq \cdot (\bk_1-\bk_2)}. 
\label{eq7}
\end{eqnarray}

%
%

Some details about the evaluation of 
$G(q)$ and 
$\Delta E$ with a zero-range force 
are recalled in the Appendix; 
finally, we write $\Delta E(\rho)/A$ as  
$\chi(\rho)\times I(\rho,\infty)$, with  
\begin{eqnarray}
&& \chi(\rho) \equiv
- \frac{3}{4 \pi^6} \frac{m k_F^7g^2}{\hbar^2 \rho}, 
\label{eq7n}
\\  
&& I(\rho,\infty) \equiv \frac{1}{15} \left( \int_0^1 u du F_1(u) + 
\int_1^{\infty} u du F_2 (u) \right)
\label{eq8}
\end{eqnarray}
where $u=q/2k_F$. When the cutoff $\Lambda$ is introduced, 
the last integral has $\Lambda/2k_F$ as upper limit and the corresponding quantity is denoted by 
$I(\rho,\Lambda)$. 
The expressions for $F_1(u)$ and $F_2(u)$ are given in the Appendix. 
The analytical expression of $I(\rho,\Lambda)$ is
\begin{widetext}
\begin{eqnarray}
&&I(\rho,\Lambda)=  \frac{1}{105} (43 -46 \ln2) -\frac{18}{35} + 
\frac{\Lambda}{35 k_F} 
+ \frac{11 \Lambda^3}{210 k_F^3} + \frac{\Lambda^5}{840 k_F^5}
+ \frac{16 \ln2}{35}+ \left( \frac{\Lambda^5}{60k_F^5}-
\frac{\Lambda^7}{1680 k_F^7} 
\right) \ln\left(\frac{\Lambda}{k_F}\right)+\left( \frac{1}{35} \right. 
\nonumber \\ 
&& \left. - \frac{\Lambda^2}{30k_F^2} 
+\frac{\Lambda^4}{48 k_F^4} - 
\frac{\Lambda^5}{120 k_F^5} + \frac{\Lambda^7}{3360 k_F^7} \right) 
\ln\left(-2 + \frac{\Lambda}{k_F}\right) - \left(  \frac{1}{35} 
- \frac{\Lambda^2}{30k_F^2}+\frac{\Lambda^4}{48 k_F^4} + 
\frac{\Lambda^5}{120 k_F^5} 
- \frac{\Lambda^7}{3360 k_F^7} \right) \ln\left(2 + \frac{\Lambda}{k_F}\right).  
\nonumber \\
\label{eq11}
\end{eqnarray}
\end{widetext}

\begin{figure}[htb]
\begin{center}
\includegraphics[width=6.5cm]{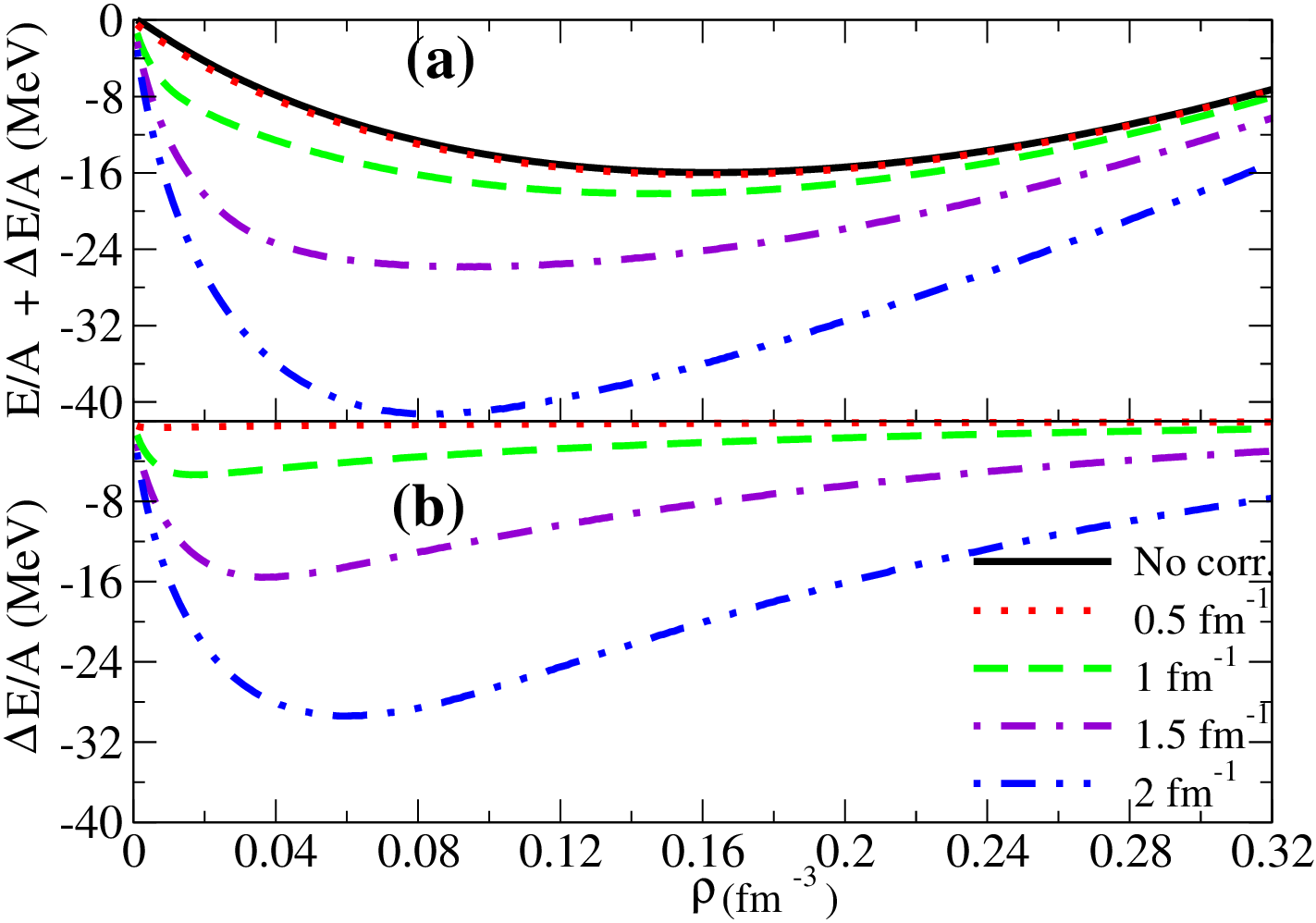}
\end{center}
\caption{\small (Color online)
 (a) $E/A + \Delta E/A$ as a function of the density 
and for 
different values of the cutoff $\Lambda$. 
The SkP mean-field EOS (solid black line) is shown for comparison. (b) Correction $\Delta E/A$ 
for different values of $\Lambda$.}
\label{fig2}
\end{figure}

In Fig. 2(a), $E/A + \Delta E /A$ is plotted for different values of 
$\Lambda$ 
and compared with the SkP mean-field 
values of $E/A$ (solid black line). The correction $\Delta E/A$ is also 
shown in Fig. 2(b). One observes that the second-order correction 
causes a shift of the saturation point to lower densities. 
The curves are calculated using the 
SkP parameters (cf. Table I). 
For $\Lambda=$ 1.5 fm$^{-1}$, the maximum correction is already comparable with the energy per 
particle at the saturation point, i.e., $\approx$ 15 MeV. 

\begin{figure}
\begin{center}
\includegraphics[width=6.5cm]{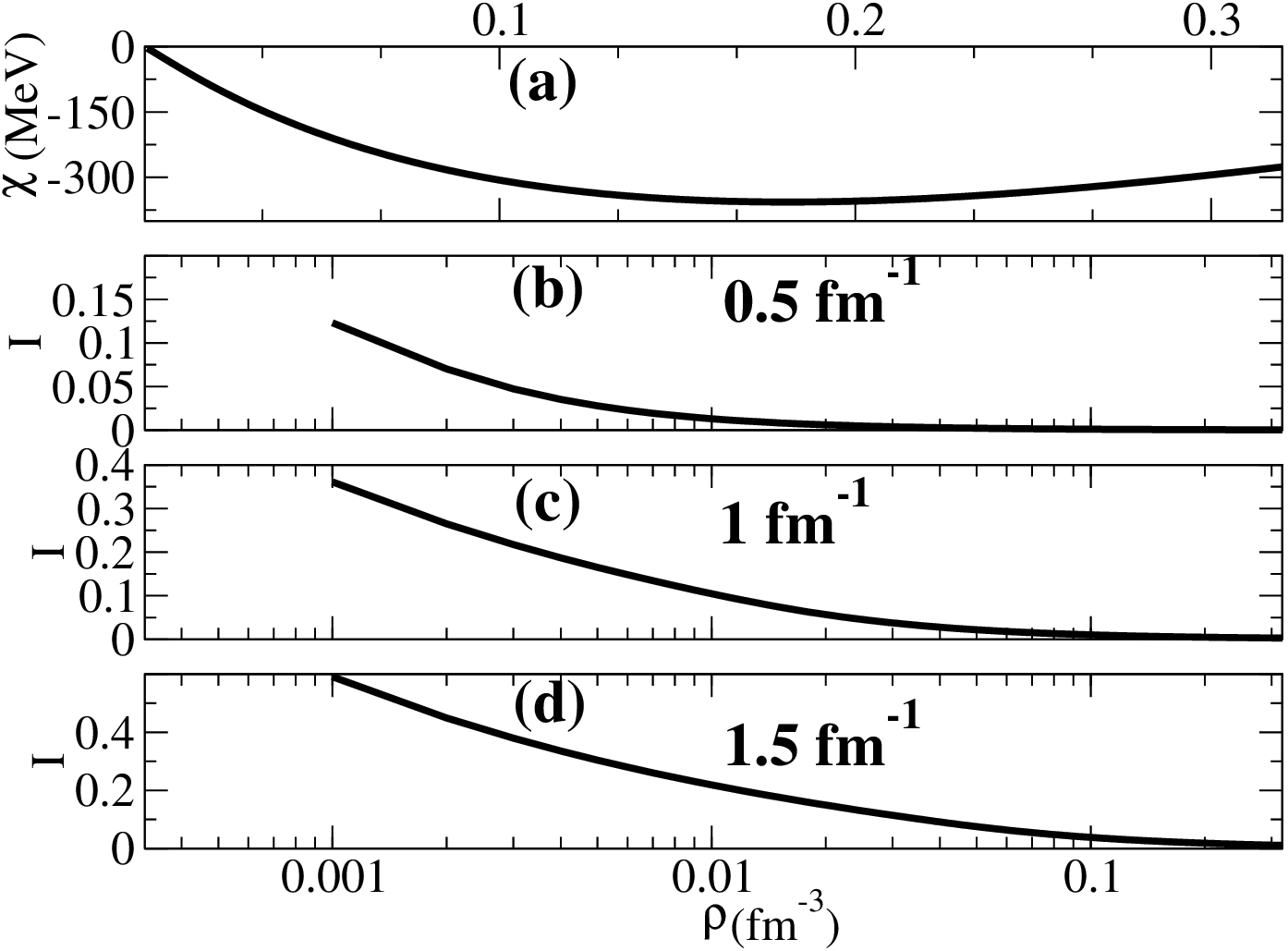}
\end{center}
\caption{\small (a) Density-dependent $\chi$ coefficient. Other panels: the values of $I(\rho,
\Lambda)$ for 
$\Lambda=$ 0.5 (b), 1 (c) and 1.5 (d) fm$^{-1}$.}
\label{fig3}
\end{figure}

To better understand the behavior of $\Delta E/A$,  
we plot  
$\chi(\rho)$ 
and $I(\rho,\Lambda)$ separately in 
Fig. \ref{fig3}.
The ultraviolet divergence is visible at all the values of the density, by comparing 
the trends in panels (b), (c) and (d). At the same time,   
one observes also a divergent behavior of $I$ for $\rho \rightarrow 0$ that is 
dictated by the upper limit 
$\Lambda/(2k_F)$
of the second integral in Eq. (\ref{eq8}). It 
can be seen in the analytical expression of Eq. (\ref{eq11}). 
$I (\rho,\Lambda)$ multiplies the coefficient $\chi (\rho)$ which 
goes to zero like $k_F^4$ 
when the density goes to zero. 
One may worry that for large values of 
$\Lambda$, 
the divergent behavior of the integral dominates when $\rho \rightarrow$ 0, 
so that no regularization is possible for very low density. However, this happens for 
values of  
$\Lambda$ which are physically meaningless, namely 
for $\Lambda\approx$ 250-300 fm$^{-1}$. 
%
  
For each value of $\Lambda$ we can perform a 
least square fit to determine a new parameter set SkP$_\Lambda$, 
such that the EOS including the second-order correction
matches rather well the one obtained with the original force
SkP at the mean-field level. This is our main result and  
it is illustrated in Fig. 4(a). 
The refitted parameters are listed in Table I (together with 
the saturation point).  
In Fig. 4(b) we display, for purely mathematical
illustration, the refit done with 
%
the extreme value of $\Lambda=$ 350 fm$^{-1}$. 


\begin{figure}[htb]
\begin{center}
\includegraphics[width=6.5cm]{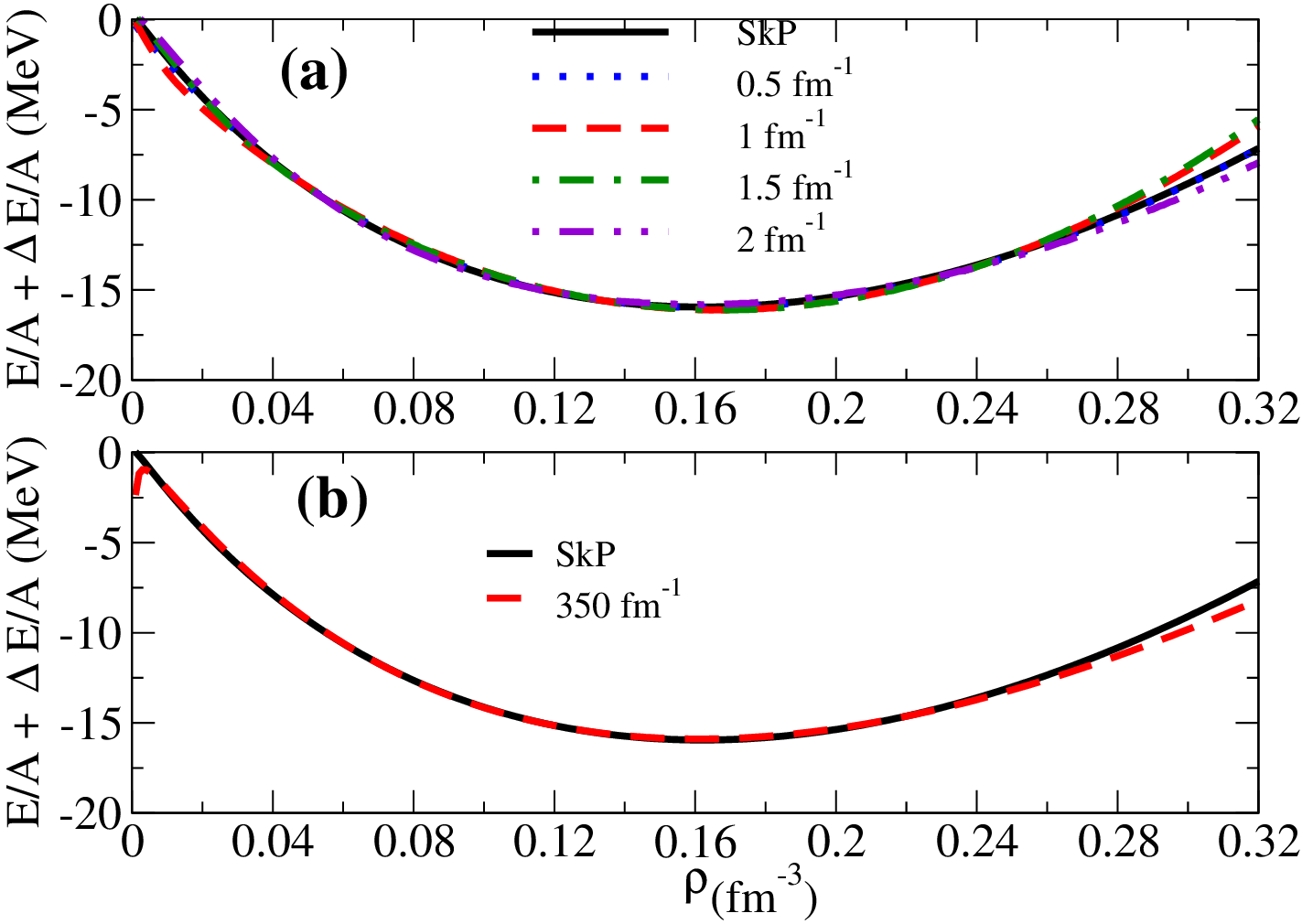}
\end{center}
\caption{\small (Color online) (a) Second-order-corrected equations of state 
compared with the reference 
equation of state (SkP at mean-field level). (b) 
Extreme case of $\Lambda=$ 350 fm$^{-1}$.}
\label{fig4}
\end{figure}

%
%
In summary we have shown that, for any physically meaningful value of the 
cutoff $\Lambda$ it is possible to find a contact interaction that 
can be used in a mean-field plus second-order corrections framework and 
can describe 
satisfactorily the empirical EOS. 
The quality of the fits, that can be judged from
Fig. 4 and the $\chi^2$-values in Table I, demonstrates that
the strategy we have outlined in a general fashion works,
in practice, for the case of symmetric nuclear matter treated
with a simplified contact force. 

\begin{widetext}

\begin{table}[!t]
\caption{From the second line, columns 2, 3 and 4: parameter sets obtained in the fits associated 
with different values of the cutoff $\Lambda$ 
compared with the original set SkP (first line). In the fifth column the $\chi^2/N$-value ($\chi^2$ divided by the number of fitted points) associated to 
each fit is shown. In columns 6 and 7 the saturation point is shown.}
\label{ta}
    \begin{ruledtabular}
      \begin{tabular}{ccccccc}
                 & $t_0$ (MeV fm$^{3}$)  & $t_3$ (MeV fm$^{3+3\alpha}$)& $\alpha$ & $\chi^2/N$  &  $\rho_0$ (fm$^{-3}$) & 
$E/A(\rho_0)$ (MeV) \\
        SkP &   -2931.70   &   18708.97 & 1/6   & &  0.16   &  -15.95  \\
        $\Lambda=$ 0.5 fm$^{-1}$ & -2352.900   & 15379.861  &  0.217   
& 0.00004 &  0.16     & -15.96  \\
        $\Lambda=$ 1 fm$^{-1}$ & -1155.580     &  9435.246   & 0.572  &  0.00142 & 0.17   & -16.11  \\
$\Lambda=$ 1.5 fm$^{-1}$ & -754.131    & 8278.251   &  1.011 & 0.00106 & 0.17 & -16.09  \\
$\Lambda=$ 2 fm$^{-1}$ & -632.653     &  5324.848  & 0.886 & 0.00192 & 0.16  & -15.82  \\
$\Lambda=$ 350 fm$^{-1}$ &  -64.904   &  360.039  &  0.425  &  0.00042 &  0.16   & -15.88  \\
      \end{tabular}
    \end{ruledtabular}

\end{table}
\end{widetext}

%
%

One can foresee 
various applications to the studies of strongly correlated fermion systems 
in all the domains of many-body physics where zero-range forces are used and where 
beyond-mean-field theories are necessary for a more accurate treatment 
of complex correlations.  
In nuclear physics, 
models beyond mean-field theories where  
ultraviolet divergences appear owing to the use of a  
zero-range interaction are: (i)  
models where multiparticle-multihole 
configurations are introduced \cite{Quentin}, (ii) 
second random-phase-approximation models 
\cite{gambacurta}, and 
(iii) models that take into account the 
coupling 
between single-particle degrees of freedom and collective vibrations 
\cite{be80,br05,co01,li07}. 
In these cases, a further development of our technique may be
envisaged. 

{\it Aknowledgments.} The authors thank Daniel Pe\~na for useful discussions. 
One of the authors (G.C.) acknowledges partial support from the Italian
research project (PRIN) named ``Many-body theory of nuclear systems and
implications on the physics of neutron stars''. 

\section*{Appendix}

$G(q)$ in 
Eq. (\ref{eq7}) can be rewritten by expressing all
wave vectors in units of $k_F$: 
\begin{eqnarray}
G(q)=k_F^4 \int_0^{\infty} d\alpha e^{-\alpha q^2} 
\int_{D_1} dk_1 e^{-\alpha \bq \cdot \bk_1 } 
\int_{D_2} dk_2 e^{\alpha \bq \cdot \bk_2 }
\label{eq15}
\end{eqnarray}
where the domains $D_i$ are $D(k_i)\equiv \left\{ k_i<1,|\bk_i+\bq|>1 
\right\}$. By 
introducing $y=\alpha q$, and the unit vector $\hat{\bq} = \bq/|q|$, 
\begin{eqnarray}
G(q)=\frac{k_F^4}{q} \int_0^{\infty} dy e^{-y q} 
\int_{D_1} dk_1 e^{-y \hat{\bq} \cdot k_1 } 
\int_{D_2} dk_2 e^{y \hat{\bq} \cdot k_2 }. 
\label{eq16}
\end{eqnarray}

1) First case: $q>2$. 
In this case, $|\bk_1+\bq|>1$ and $|\bk_2-\bq|>1$
are satisfied when $k_1<1$ and $k_2<1$. 
Eq. (\ref{eq16}) leads to
\begin{eqnarray}
G_1(q)=\frac{k_F^4}{q} \int_0^{\infty} dy e^{-y q}
\left[ \frac{2 \pi}{y^3} \left( e^y (y-1) + e^{-y} (y+1) \right) \right]^2. 
\label{eq16n}
\end{eqnarray}

2) Second case: $0<q<2$.
One can apply the technique of 
Ref. \cite{dubois}, with 
this 
change of variables:
\begin{eqnarray}
\int_{D(p)} dp f(p,q) = q \int_0^{2\pi} d\Phi \int_0^1 d \alpha 
\int_{q\alpha/2}^1 x dx f(n-\alpha q,q), 
\label{eq17}
\end{eqnarray}
where $x=\hat{q} \cdot n$ and $n$ is a unit vector $n=p+\alpha q$. 
In this case, Eq. (\ref{eq16}) becomes 
\begin{eqnarray}
&& G_2(q)=\frac{k_F^4}{q} \int_0^{\infty} dy e^{-y q}
\left[ \frac{2 \pi}{y^3} \left( e^{-y} (y+1) - e^{y(q-1)} (y+1) 
\right. \right. \nonumber \\
&& \left. \left. + 
qy e^{\frac{yq}{2}} \right) \right]^2. 
\label{eq17n}
\end{eqnarray}
The energy correction can be written as
\begin{eqnarray}
\Delta E = 4 \pi C k_F^3 \left( \int_0^2 q^2 dq v^2 G_1(q) + \int_2^{\infty} q^2 dq v^2 G_2(q) \right).
\label{eq18}
\end{eqnarray}
With the 
change 
$u=q/2$ 
we obtain $\Delta E(\rho)/A$ as  
$\chi(\rho)\times I(\rho,\infty)$ [cf.  
Eqs. (\ref{eq7n}) and (\ref{eq8})], with 
\begin{eqnarray}
F_1(u) &=& \left(4+ \frac{15}{2} u -5u^3 + \frac{3}{2} u^5 \right) log(1+u)
\nonumber \\
&& + 
\left(4- \frac{15}{2} u +5u^3 - \frac{3}{2} u^5 \right) log(1-u)
 \nonumber \\
&+& 29 u^2 -3 u^4 -40 u^2 log2;
\label{b1}
\end{eqnarray}
\begin{eqnarray}
F_2(u) &=& \left(4- 20 u^2 -20 u^3 + 4 u^5 \right) log(1+u) \nonumber \\ &&+ 
\left(-4+ 20 u^2 -20 u^3 + 4 u^5 \right) log(u-1)
 \nonumber \\
&+& 22 u +4 u^3 +(40 u^3 -8 u^5) logu.
\label{b2}
\end{eqnarray}

\providecommand{\noopsort}[1]{}\providecommand{\singleletter}[1]{#1}%

\end{document}